\documentclass[twocolumn,english,aps,prb,floatfix,showpacs]{revtex4}
\usepackage{graphicx}
\usepackage{amssymb}

\makeatletter

\usepackage{babel}

\usepackage{babel}

\usepackage{babel}
\makeatother
\begin{document}

\title{Tailoring Graphene with Metals on Top}

\author{B. Uchoa$^{1}$, C.-Y. Lin$^{2}$, and A. H. Castro Neto$^{1}$}

\affiliation{$^{1}$ Physics Department, Boston University, 590 Commonwealth Ave.,
Boston, MA 02215}

\affiliation{$^{2}$ IBM Almaden Research Center, San Jose, CA 95120}

\begin{abstract}
We study the effects of metallic doping on the electronic properties of graphene
using density functional theory in the local density approximation in the presence
of a local charging energy (LDA+U). The electronic properties are sensitive to whether graphene is
doped with alkali or transition metals.  We estimate the 
the charge transfer from a single layer of Potassium on top of graphene in terms of the local charging energy of the graphene sheet. The coating of graphene with a {\it non-magnetic} layer of Palladium, on the other hand, can lead to a magnetic instability in coated graphene due to the hybridization between the transition-metal and the carbon orbitals.
\end{abstract}


\maketitle

\section{Introduction}
Graphene, a two dimensional (2D) allotrope of carbon on a honeycomb lattice,
is characterized by 
elementary electronic excitations described in terms of Dirac fermions \cite{Novoselov},
a solid state realization of a relativistic system \cite{pw}.
The observation of theoretically predicted \cite{sharapov,Peres}
anomalous plateaus in the quantum
Hall effect \cite{Novoselov2,Zhang} and of a universal
minimum of conductivity \cite{Novoselov2} attracted a lot of interest. 
In the absence of doping, 
graphene behaves as an unusual metal with low density of states \cite{geim}. 
Because the linear band spectrum
is a robust feature of the honeycomb lattice, the excitations in graphene
are particles with zero effective mass that propagate coherently 
very large distances disregarding the amount of impurities, allowing
graphene to sustain supercurrents \cite{Heershe}. By applying a bias
voltage, the carrier density of graphene can be controlled by
electric field effect allowing for many practical applications
ranging from the production
of electronic lenses \cite{Cheianov} to the fabrication of semiconductors
with a tunable gap in bilayers \cite{Castro}. 

One of the roads still unexplored in the material science of 
graphene is the tailoring of its electronic properties by chemically adsorption
of metallic atoms. The electronic properties that result from adsorption
depend strongly on the ionic and/or covalent character of the bonds formed
between carbon and the metal. Alkaline metals are good electron donors
because of the strong ionic character of their bonding, 
increasing dramatically the number of charge carriers in graphene \cite{ohta}.
At the same time, the metallic bands can be strongly affected by the
presence of the graphene lattice, as observed in some graphite
intercalated compounds (GIC) \cite{Csanyi}. On theoretical grounds,
both effects can produce superconductivity in coated graphene \cite{Uchoa}.
Transition metals by their turn tend to make strong covalent bonds.
Because of the strong electron-electron interactions in these materials,
they are
more susceptible to induce magnetic instabilities. The tuning of magnetism
in coated graphene can open other routes to spintronics through new
spin-valve devices \cite{tati}. Similar effects have been studied
in the context of carbon nanotubes \cite{solange}.

We study the effect of metallic coating of graphene using
two different metal atoms, potassium (K), and palladium (Pd). 
Among alkaline metals, K atoms have
a particularly good matching with the graphene lattice and their adsorption
on graphite surfaces has been a topic of extensive research \cite{Caragiu}.
We address the problem of charge transfer between K and graphene, 
and discuss the nature of the K conduction band in coated graphene.
Although there are many transition atoms that interact
strongly with carbon \cite{solange2}, Pd has very polarizable bands 
and hence, is a natural candidate for the generation of magnetism 
in coated graphene. 

\section{Numerical calculation}

We investigate the electronic properties of coated graphene from a
band structure calculation based in density functional
theory (DFT) in local density approximation (LDA). In order to study
a single-layer of graphene, we repeat graphene layers periodically with 33.6 {\AA }
of vacuum separation. The electronic structure is calculated using
the all-electron full-potential linearized augmented plane wave (FLAPW)
method \cite{wien2k} with corrections to the exchange-correlation
potential calculated in the generalized gradient approximation (GGA)
\cite{perdew}. To further
take into account the local interactions, we go beyond GGA assigning
a phenomenological parameter $U$ (LDA+U) \cite{Anisimov,Liechtenstein},
which corresponds to the effective local potential between two electrons
in the same localized orbital. 

\subsection{K coated graphene}

In K coated graphene, K atoms sit on top of the hexagons around 2.7
{\AA } away from the graphene plane \cite{Lamoen}. Since K is much
larger than carbon, the maximum coverage is achieved at one K atom
per $8$ carbons, KC$_{8}$, what corresponds to one monolayer. In
this concentration, we relax the K-graphene distance to 2.68
{\AA } within GGA after the minimization of the interatomic forces
is done. Since the C-C bonds are much stronger than the bonds of C
with the metal, the effect of the relaxation of the graphene lattice
due to the coating is a very small effect \cite{Ancilotto}. We use
the lattice parameter of pure graphene, $a=1.42$ {\AA}, in the calculations. 

\begin{figure}
\begin{center}\includegraphics[%
  scale=0.41]{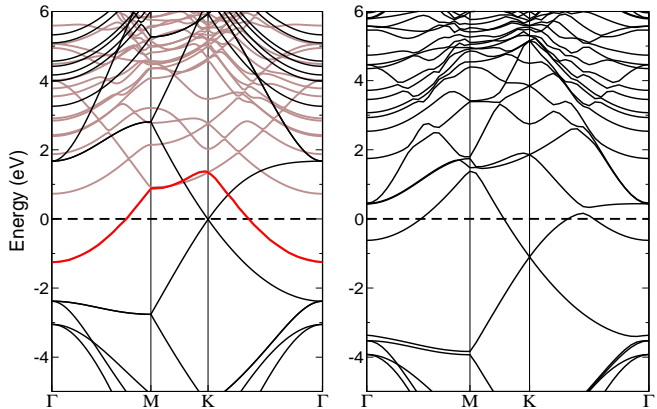}\hspace{0cm}
\end{center}

\caption{{\small (color online) On the left:  superposition of the bands for 
one monolayer of K (red and brown lines) and pure graphene (black lines) 
in the unit cell of KC$_8$. On the right: band plot of KC$_8$ in GGA  ($U=0$).}}
\end{figure}

When K is deposited on top of graphene, part of the electrons in the
4$s$-band are transfered to the $\pi^{*}$-band of C to compensate
for the difference in electronegativities. In this process, the $\pi$-bands
suffer a rigid shift, generating a pocket of electrons with a finite
density of charge carriers in the graphene plane. At the same time,
there is a significant redistribution of charges within the K states
which accompanies the formation of K-C bonds \cite{Ancilotto,Li}.
In the plot on Fig. 1 (left), we show the bands of graphene in the
expanded unit cell of KC$_{8}$ superimposed with the bands of a free standing
K monolayer. The 4$s$ band of K shown in red
 forms a nearly circular Fermi surface around the center
of the Brillouin zone (BZ) at the $\Gamma$ point, while the $\pi$-bands
of pure graphene cross the Fermi level at the corners of the BZ
(K point). After a careful comparison between the high energy bands
of KC$_{8}$ with the K and graphene bands in separate, we are able
to find a nearly one-to-one correspondence in the energy range 0-6 eV 
between the empty bands 
of KC$_{8}$, shown in Fig. 1 (right), and the empty bands of the free K monolayer. 
However, we find no trace of
the graphene nearly free electron (NFE) bands in KC$_{8}$, which
in pure graphene start from $\sim$ 3 eV above the Fermi level, as shown in Fig. 1
 \cite{Posternak}. This suggests that the interstitial NFE
bands drop to the Fermi level and hybridize with the 4$s$ band of
K in a similar way to previous band structure analysis on GIC \cite{Csanyi}
and carbon nanotubes \cite{Miyamoto}. The strong downshift of these
high energy bands is an electrostatic response of the graphene
NFE bands to the potential induced by the background of positive charge
of the K ions \cite{Margine}. 
This picture indicates that
the charge of this system is distributed in the graphene layer and
also in the interstitial space of the KC$_{8}$ bilayer.

\begin{figure}[b]
\begin{center}\includegraphics[%
  scale=0.33]{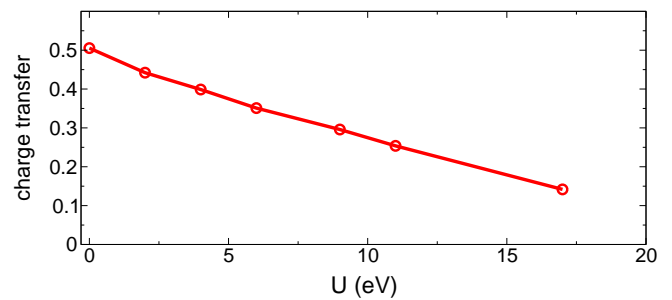}\vspace{-0.3cm}\end{center}

\caption{{\small  Charge transfer
$\delta Q$ from K to graphene in KC$_{8}$ (in electrons per K) as a function
of the effective Hubbard  $U$ in the C $p_{z}$ orbitals. The charge
transfer for $U=0$ has been calculated in GGA, for comparison with Ref. \cite{Lou}.}}
\end{figure}

The amount of charge transfer to graphene in KC$_{8}$ is still an
open question in the literature and has been studied for many years 
in K deposited
graphite surfaces \cite{Caragiu}. Previous LDA calculations with
specific assumptions on the pseudopotentials have predicted a charge
transfer of $\sim0.17\: e$ per K \cite{Lamoen,Ancilotto} on KC$_{8}$
and $\sim0.1\: e$ per K for one monolayer of K on top o a stack of
graphite layers \cite{Rytkonen, Pivetta}. On the other hand, a recent all-electron
GGA cluster method calculation on KC$_{8}$ has predicted a substantially
larger charge transfer of $0.46$ $e$ per K \cite{Lou}. We remark
however that because C orbitals are poorly screened by K, the local
Hubbard potential in the carbon $p_{z}$ orbitals is an important
input which modifies the difference of electronegativities between
graphene and the metal. Indeed, we observe within the LDA+U framework
that the total amount of charge transfer is very sensitive to $U$. 
To estimate the charge transfer, we extract the energy shift of the K 4$s$ band
(red curve in Fig. 1) when the K layer is deposited on top of graphene. 
For $U=0$, our DFT calculation predicts a charge transfer of $\delta Q= 0.51\, e$
per K. As $ U $ increases the charge transfer is linearly suppressed and
eventually extrapolates to zero for $U\sim25$ eV (see Fig.
2). The hopping energy of the electrons in graphene is $t\sim3\,$eV. As in fullerenes 
\cite{Gunnarsson}, 
if a Hubbard $U\sim3t$ can be taken as a reasonable estimate for the order of magnitude 
of $U$ in graphene, then the LDA+U calculation may lower considerably the
amount of charge transfer predicted by a pure GGA
calculation \cite{Lou}. On the other hand, since the 
LDA+U method does not take polarization effects into account,
the local charging energy $U$ can be additionally screened by the pocket 
of electrons in graphene. Our effective $U$ is a screened local charging 
energy in the graphene layer.  

The electronic density transfered to the graphene layer is $\delta Q/A_{\mathrm K}$ 
where $A_{\mathrm K}=6\sqrt{3}a^2$ is the area of the K unit cell on top of graphene. 
For a small charge transfer $\delta Q$,  the shift in the chemical potential 
of the $\pi$-bands is 
\begin{eqnarray}
\delta \mu=(v_{F}/a)\sqrt{\pi/(6\sqrt{3})\,\delta Q}\,\sim2.3\,\sqrt{\delta Q} \,
{\rm eV} 
\end{eqnarray}
where $v_{F}=6$eV {\AA} is the $\pi^*$-band velocity around the K point. 
Hence, we see that a very small charge transfer
of $\delta Q\sim0.01\, e$ per K is already enough to exceed the maximum
chemical shift achievable by the application of a bias voltage in graphene
($\approx 0.2$ eV \cite{Novoselov}).  
For a charge transfer larger than $\sim0.2\,e$ per K, the shift 
of the $\pi$-bands is above $1$ eV and reaches $\delta \mu \sim 1.2$ eV for $U=0$, 
when the charge transfer is maximum (see Fig. 1).  
A $\pi$-band shift of the same order has been very recently measured with angle resolved 
spectroscopy (ARPES) for K coated graphene (KC$_8$) on a SiC substrate \cite{McChesney}.
In this experiment, the charge transfered from the substrate shifts  the $\pi$ bands of pure graphene 
in $0.45$ eV, which corresponds to $\sim0.04\,e$ per unit cell of KC$_8$. The total $\pi$-band shift 
was estimated in $\sim 1.2$ eV, indicating that each K atom transfers the order of $0.5\,e$ to graphene. This result contrasts with a recent photoemission experiment for a K monolayer adsorbed on the surface of graphite, which estimated the charge transfer in $\sim0.1\,e$ per K\cite{Algdal}. According to our DFT results, these experiments suggest that the local charging energy of graphene is strongly screened by the presence of the SiC substrate (see Fig. 2). 

In K intercalated graphite, previous DFT pseudopotential  calculations  have reported charge transfers larger than 0.7 $e$ per K \cite{DiVicenzo, Hartwigsen}. These results may indicate that the character of the K/C bonding is modified by intercalation \cite{Hartwigsen}. A more conclusive comparison between GIC with the present results, however, requires a closer examination of the intercalated case through a similar DFT method. 

The size control of the charge pockets in graphene can be achieved by dilution of the K coverage. For dilution less than $0.9$ of a monolayer, however, K does not form a uniform metallic lattice due to clustering,  leading to insulating behavior \cite{Caragiu}. 

\subsection{Pd coating case}

Because of the low coordination number, the $d$ orbitals of transition
metals can be very localized in systems of low dimensionality. Among
4$d$ transition metals, bulk paramagnetic elements like Ru, Rh, and
Pd exhibit magnetism on surfaces and in nanosystems \cite{Pfandzelter,Goldoni2,Sampedro}.
In particular, Pd atoms are not intrinsically magnetic because of
the closed shell configuration of the 4$d$ orbitals (4$d^{10}$5$s^{0}$).
In bulk, the $s$ and $d$ bands of Pd hybridize (4$d^{10-\epsilon}$5$s^{\epsilon}$),
and Pd exhibits strong Pauli paramagnetism with a high magnetic susceptibility.
Due to the high density of states near the Fermi surface, bulk fcc
Pd is close to a ferromagnetic instability. Expanding the fcc lattice
in 5\%, the Stoner criterion is satisfied and Pd becomes an itinerant
ferromagnet due to the enhancement of the $s-d$ band hybridization
\cite{Chen1}. 

On top of graphite, Pd atoms do not form a uniform metallic coating
but large clusters \cite{Egelholf}. It is not clear to what extent
Pd atoms can grow laterally to form islands as Ru \cite{Pfandzelter}
and Rh \cite{Goldoni} or if they agglomerate to form clusters of
a few layers thick. If we assume that
the Pd atoms inside the islands form a low temperature commensurate
structure with the graphene lattice with the same periodicity as in
Rh adsorbed on the surface of graphite \cite{Goldoni}, the Pd atoms
sit at the center of the hexagons of graphene separated by a
distance of $\sqrt{3}a\sim2.46${\AA} from the next Pd neighbor.
In this case, the Pd-graphene equilibrium distance is $3.35$ {\AA},
after we minimize the interatomic forces between the Pd and graphene
layers. This configuration represents a lateral compression of $\sim$7\%
with respect to the bond length of Pd in a square lattice \cite{Nautyial},
and a very small magnetization is expected as a result of the weak
hybridization of the bands \cite{chen}. 

\begin{figure}
\begin{center}

\includegraphics[scale=0.45]{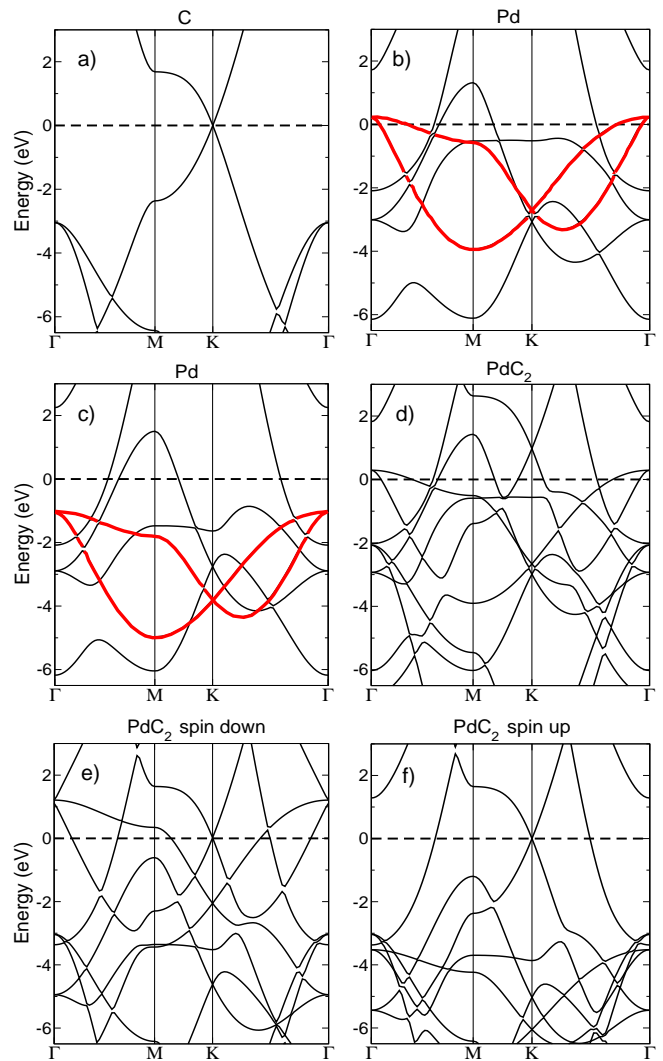}

\end{center}

\caption{{\small (color online) Band plots for (a) graphene, (b) a free standing
monolayer of Pd for $U=0$ and (c) free Pd monolayer for $U=11$ eV.
In plots (d), (e) and (f) we display the bands of PdC$_{2}$ for:
(d) $U=0$,  (e) $U=$11 eV in Pd for spin down and (f) $U=11$eV in Pd for
spin up. }}
\end{figure}

Nevertheless, the splitting of the bands due to the hybridization of Pd
and graphene orbitals at finite $U$ can produce a strong enhancement
of the density of states at the Fermi surface. We show that the Pd bands in 
coated graphene can strongly polarize through a Stoner transition. 
We focus in the Pd coating to illustrate 
the effect of the metal-carbon hybridization in the production of 
itinerant ferromagnetism. Since the orbitals in the transition metal monolayer
are more localized than in graphene, we assign in first approximation 
a single effective parameter $U$ to the metallic bands in our LDA+U calculation.

In Fig. 3(b) we show the spectrum of a free standing monolayer of Pd
in the unit cell of graphene. The bands in red can be associated at the $\Gamma$
point (the center of the BZ) to the two degenerated bands $4d_{xz}$
and $4d_{yz}$ of the free Pd atom. The two bands that cross the $\Gamma$ 
point at $\sim -3$ eV 
can be associated near $\Gamma$ to the degenerated bands $4d_{xy}$
and $4d_{x^{2}-y^{2}}$ of free Pd, while the 5$s$-band crosses the
$\Gamma$ point at $\sim-2$eV. Finally, the high energy band which
crosses the $\Gamma$ point at $\sim6$ eV below the Fermi level corresponds
to the $4d_{z^{2}}$ orbital of free Pd around the center of the BZ.

When we turn on the local potential to $U=11$ eV in the free Pd monolayer
{[}see Fig. 3(c){]}, we clearly see that the bands in red 
suffer a rigid downshift of $\sim1$ eV, indicating the presence
of localized states in these bands. The band which is nearly flat between 
points K and M in Fig. 3(b) 
shifts in $\sim1$ eV around the border of the BZ (points K and M),
where the effects of the lattice symmetry are stronger, and remains
unaltered at the center of the BZ, where the $d_{x^{2}-y^{2}}$, $d_{xy}$
orbital symmetries dominate, suggesting that the states around $\Gamma$
in this band are more delocalized along the plane of the Pd monolayer.
The other two $4d$ bands and the 5$s$ band are almost insensitive
to the local potential and are delocalized bands. We do not see any
trace of magnetization in the free standing Pd monolayer for $U$ in the range of
$0-12$ eV. 

\begin{figure}
\begin{center}\includegraphics[%
  scale=0.31]{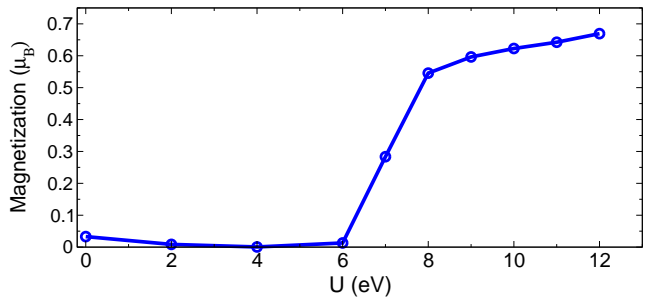}\hspace{0cm}

\caption{{\small Magnetization of Pd coated graphene, PdC$_2$, (in Bohr magneton) {\it vs.} local potential $U$ in the Pd orbitals.}}
\end{center}

\end{figure}

In the sequence of plots, in Fig. 3(d), (e) and (f), we put the Pd
monolayer on top of graphene for $U=0$ and then we set $U$ finite. Because graphene
is more electronegative than Pd, the $\pi$ band is shifted up in $\sim 1$ eV for $U=0$.
In this case, the $d$ orbitals of Pd strongly hybridize with the graphene $p_{z}$ orbitals 
but the Stoner condition is not satisfied and the system does not magnetize. 
For $U= 11$ eV, however, the $\pi$-bands nearly recover charge neutrality and the 
redistribution of charge among the hybridized $d$ orbitals due to the local charging energy enhances dramatically the density of states in the localized bands of Pd at the Fermi surface. 
In this situation, the Pd bands strongly polarize, as displayed in Fig. 3 (e)
and (f), and we observe a total magnetization of 0.64 $\mu_B$ in the Pd unit cell (see Fig. 4), indicating a ferromagnetic instability. The strength of the magnetization is very sensitive
to the specific value of the assigned local potential for $U\gtrsim 6$ eV, where 
the Stoner condition is satisfied and the system has a ferromagnetic 
phase transition, as shown in Fig. 4.

\section{Summary and conclusion}

We have explored the electronic properties of graphene
coated with alkaline and transition metals from a band structure calculation
in LDA+U. We propose that graphene can be tailored either by the control of 
the number of charge carriers,  
which can be tuned by coating graphene with a controlled coverage of alkaline
metals, such as K, or by inducing a ferromagnetic instability in graphene through the
coating with a transition metal. 

We addressed the problem of the charge transfer between 
one monolayer of K and graphene, emphasizing the role of the local charging energy 
in the graphene layer. Our DFT results in coated graphene suggest that the metallic band 
of K hybridizes with the carbon NFE band in a similar way as previously reported in GIC and 
carbon-nanotubes. We emphasize that the problem of coating is conceptually different 
from the adsorption of isolated alkaline adatoms in graphene, where the metallic orbitals strongly localize around the impurities. In particular, the estimation of the charge transfer from each K impurity in the dilute phase depends on specific definitions of the atomic charge \cite{Valencia}. In graphene coated with a uniform monolayer of K, the K $s$ electrons delocalize and the charge transfer can be accurately estimated by the energy shift of the bands at the Fermi level, as shown in Fig. 1. 

The amount of charge transfered from K within the LDA+U framework is sensitive to the effective 
local charging energy assigned to the carbon orbitals. For $U=0$, we find a maximum
charge transfer of 0.51 $e$ per K, which is linearly suppressed with the increase of the 
effective charging energy $U$. A comparison of our DFT calculation with recent spectroscopy 
and photoemission experiments suggest that different substrates can give rise to very different 
values of $U$ in graphene.  

For $4d$ transition metals such as Pd, we showed
that the hybridization of the graphene $p_{z}$ orbital with the localized $d$ orbitals 
at finite $U$ can produce strong itinerant magnetism in coated graphene. The onset of 
ferromagnetism in Pd coated graphene is identified through a Stoner transition for $U\gtrsim6$ 
eV, where the local bands of Pd polarize.  

\section{Acknowledgments}

We thank S. Fagan for illuminating discussions. B.U. acknowledges
CNPq, Brazil, for the support under the grant No. 201007/2005-3. 
A.H.C.N. was supported through NSF grant DMR-0343790.

\end{document}